\documentstyle[12pt,epsfig]{article}
\date{}
\begin{document}

{\large \sf
\title{
{\normalsize
\begin{flushright}
~~~
\end{flushright}}
{\vspace{2cm} {\LARGE \sf
Jarlskog Invariant of \\
the Neutrino Mapping Matrix} \vspace{3cm}} }

{\large \sf
\author{
{\large \sf
R. Friedberg$^1$ and  T. D. Lee$^{1,~2}$}\\
{\normalsize \it 1. Physics Department, Columbia University}\\
{\normalsize \it New York, NY 10027, U.S.A.}\\
{\normalsize \it 2. China Center of Advanced Science and Technology (CCAST/World Lab.)}\\
{\normalsize \it P.O. Box 8730, Beijing 100080, China}\\
} \maketitle

\begin{abstract}

{\normalsize \sf

The Jarlskog Invariant $J_{\nu-map}$ of the neutrino mapping matrix
is calculated based on a phenomenological model which relates the
smallness of light lepton masses $m_e$ and $m_1$ (of $\nu_1$) with
the smallness of $T$ violation. For small $T$ violating phase
$\chi_l$ in the lepton sector, $J_{\nu-map}$ is proportional to
$\chi_l$, but $m_e$ and $m_1$ are proportional to $\chi_l^2$. This
leads to $ J_{\nu-map}  \cong \frac{1}{6}\sqrt{\frac{m_e}{m_\mu}}+O
\bigg(\sqrt{\frac{m_em_\mu}{m_\tau^2}}\bigg)+O
\bigg(\sqrt{\frac{m_1m_2}{m_3^2}}\bigg)$. Assuming
$\sqrt{\frac{m_1m_2}{m_3^2}}<<\sqrt{\frac{m_e}{m_\mu}}$, we find
$J_{\nu-map}\cong 1.16\times 10^{-2}$, consistent with the present
experimental data.

 }
\end{abstract}

{\normalsize \sf PACS{:~~14.60.Pq,~~11.30.Er}}

\vspace{1cm}


\newpage

\section*{\Large \sf  1. Introduction}
\setcounter{section}{1} \setcounter{equation}{0}

In a recent paper[1], we proposed a phenomenological model in which
the smallness of light quark masses $m_d$ and $m_u$ are related to
the smallness of $T$ violation. Thus, when the relevant $T$
violating phase $\chi_q$ in the quark sector is nonzero but small,
$m_d$ and $m_u$ are proportional to $\chi_q^2$. On the other hand,
the Jarlskog invariant $J_{CKM}$ of the CKM matrix depends linearly
on $\chi_q$, which leads to a relation
$$
J_{CKM}  \cong
A\lambda^3\bigg[\sqrt{\frac{m_dm_s}{m_b^2}}
+O\bigg(\sqrt{\frac{m_um_c}{m_t^2}}\bigg)\bigg]
$$
$$
\cong 3\times 10^{-5}\eqno(1.1)
$$
where $A\cong 0.818$ and $\lambda\cong 0.227$ are the Wolfenstein
parameters. In this paper, we continue the model analysis for the
lepton sector. As we shall see, in the small $T$ violation limit the
corresponding Jarlskog invariant $J_{\nu-map}$ of the neutrino
mapping matrix is related to the lepton masses by
$$
J_{\nu-map}  \cong \frac{1}{6}\sqrt{\frac{m_e}{m_\mu}}+O
\bigg(\sqrt{\frac{m_em_\mu}{m_\tau^2}}\bigg)+O
\bigg(\sqrt{\frac{m_1m_2}{m_3^2}}\bigg)\eqno(1.2)
$$
where $m_1$, $m_2$ and $m_3$ are masses of $\nu_1$, $\nu_2$ and
$\nu_3$. Assuming
$$
\frac{m_1m_2}{m_3^2}<<\frac{m_e}{m_\mu},\eqno(1.3)
$$
we find
$$
J_{\nu map}\cong \frac{1}{6}\sqrt{\frac{m_e}{m_\mu}} \cong
1.16\times 10^{-2}\eqno(1.4)
$$
consistent with the present ($1~\sigma$- deviation) experimental
limit[2]
$$
J_{\nu-map} \leq 4.2 \times 10^{-2}.\eqno(1.5)
$$
Thus, an improvement of a factor $4$ of the experimental accuracy
could yield a meaningful test of the model.

The underlying reasoning of our analysis is a spontaneous $T$
violation field theory model [3,4], in which the Higgs field
responsible for $T$ violation belongs to the same family of scalar
fields that generates masses for light quarks and light leptons.
However, in this paper and Ref.~1, we restrict our discussions only
to a phenomenological analysis of the relevant mass matrices. As in
Ref.~1, let $l_i(\downarrow)$ and $l_i(\uparrow)$ be the
hypothetical lepton states "diagonal" in $W^\pm$ transitions:
$$
l_i(\downarrow)\rightleftharpoons l_i(\uparrow) + W^-
$$
$${\sf and}~~~~~~~~~~~~~~~~~~~~~~~~~~~~~~~~~~~~~~~~~~~~~~~~~~~~~~\eqno(1.6)$$
$$
l_i(\uparrow)\rightleftharpoons l_i(\downarrow) + W^+
$$
with $i=1,~2,~3$. Their electric charges in units of $e$ are $-1$
for $l_i(\downarrow)$, and $0$ for $l_i(\uparrow)$. For each of
these triplets, there exists a $3\times 3$ mass matrix $M(l)$ with
the corresponding mass operator ${\cal M}(l)$ given by
$$
{\cal
M}(l_\downarrow)=\bigg(\bar{l}_1(\downarrow),~
\bar{l}_2(\downarrow),~\bar{l}_3(\downarrow)\bigg)
M(l_\downarrow) \left(
\begin{array}{r}
l_1(\downarrow)\\
l_2(\downarrow)\\
l_3(\downarrow)
\end{array}\right), \eqno(1.7)
$$
and
$$
{\cal
M}(l_\uparrow)=\bigg(\bar{l}_1(\uparrow),~
\bar{l}_2(\uparrow),~\bar{l}_3(\uparrow)\bigg)
M(l_\uparrow) \left(
\begin{array}{r}
l_1(\uparrow)\\
l_2(\uparrow)\\
l_3(\uparrow)
\end{array}\right)\eqno(1.8)
$$
in which the related Dirac field operator $\psi$ and its Hermitian
conjugate $\psi^\dag$ are given by
$$\psi_{i\downarrow}=l_i(\downarrow),~~
\psi^\dag_{i\downarrow}\gamma_4=\bar{l}_i(\downarrow)\eqno(1.9)
$$
and likewise for the $\uparrow$ sector.

In the zeroth approximation of $T$ invariance, $M(l_\downarrow)$ and
$M(l_\uparrow)$ are both real. At the same time, we assume that the
mass operator ${\cal M}(l_\downarrow)$ satisfies a hidden symmetry
$$
l_1(\downarrow)\rightarrow l_1(\downarrow)+z,
~~l_2(\downarrow)\rightarrow l_2(\downarrow)+\eta_\downarrow z
~~{\sf and}~~ l_3(\downarrow)\rightarrow
l_3(\downarrow)+\xi_\downarrow\eta_\downarrow z \eqno(1.10)
$$
where $\xi_\downarrow,~\eta_\downarrow$ are $c$-numbers and $z$ is a
space-time independent constant element, anticommuting with the
Dirac field operators. A similar symmetry is assumed for the
$\uparrow$ sector. As shown in Ref.~1, in each sector $\downarrow$
and $\uparrow$, this symmetry yields a zero mass particle state
(i.e., $e$ and $\nu_1$). Thus, we may write the zeroth approximation
of ${\cal M}(l_\downarrow)$ and ${\cal M}(l_\uparrow)$ as
$$
{\cal M}_0(l_\downarrow) = \alpha_\downarrow | l_3(\downarrow) -
\xi_\downarrow l_2(\downarrow)|^2 + \beta_\downarrow |
l_2(\downarrow) - \eta_\downarrow l_1(\downarrow)|^2
+\gamma_\downarrow |l_1(\downarrow) - \zeta_\downarrow
l_3(\downarrow)|^2\eqno(1.11)
$$
and
$$
{\cal M}_0(l_\uparrow) = \alpha_\uparrow | l_3(\uparrow) -
\xi_\uparrow l_2(\uparrow)|^2 + \beta_\uparrow | l_2(\uparrow) -
\eta_\uparrow l_1(\uparrow)|^2 +\gamma_\uparrow |l_1(\uparrow) -
\zeta_\uparrow l_3(\uparrow)|^2\eqno(1.12)
$$
with the 12 parameters $\alpha_\downarrow,~\xi_\downarrow,\cdots$
all real. The symmetry (1.10) for the $\downarrow$ and $\uparrow$
sectors requires
$$
\xi_\downarrow\eta_\downarrow\zeta_\downarrow=1\eqno(1.13)
$$
and
$$
\xi_\uparrow\eta_\uparrow\zeta_\uparrow=1.\eqno(1.14)
$$
Consequently, in the approximation of $T$ invariance the model
contains
$$
2\times (6-1)=10\eqno(1.15)
$$
real parameters. On the other hand, in the same approximation, there
are 4 nonzero masses of $\mu,~\tau,~ \nu_2$ and $\nu_3$, In addition
the $\nu$-mapping matrix is real, specified by $3$ parameters. Thus,
the total number of observables are
$$
4+3=7.\eqno(1.16)
$$
The difference
$$
10-7=3\eqno(1.17)
$$
gives the three "gauge" degrees of freedom that are needed to
specify the orientation of the $3$-dimensional hidden frame
$\Sigma_l$ characterized by its ortho-normal axes
$\hat{l}_1,~\hat{l}_2$ and $\hat{l}_3$, with
$\hat{l}_i=\hat{l}_i(\downarrow)=\hat{l}_i(\uparrow)$ and
$i=1,~2,~3$.

In Ref.~1, two of these "gauge" degrees of freedom are eliminated by
requiring
$$\beta_\downarrow=\gamma_\downarrow\zeta_\downarrow^2~~ {\sf and}~~
\beta_\uparrow=\gamma_\uparrow\zeta_\uparrow^2.\eqno(1.18)
$$
As we shall see, these two conditions have a simple geometrical
interpretation. Let $\Sigma_\nu$ be the reference frame with its
base vectors defined by the physical state-vectors
$\hat{\nu}_1,~\hat{\nu}_2$ and $\hat{\nu}_3$. Likewise, let
$\Sigma_e$ be the corresponding reference frame whose base vectors
are the physical state-vectors $\hat{e},~\hat{\mu}$ and
$\hat{\tau}$. As we shall see, the two conditions in (1.18) are
equivalent to choosing the orientation of the $\hat{l}_1$ axis in
the hidden frame $\Sigma_l$ to be perpendicular to $\hat{\nu}_3$ and
$\hat{\tau}$; i.e.,
$$
\hat{l}_1\parallel \hat{\nu}_3\times\hat{\tau}.\eqno(1.19)
$$
To establish this relation, we follow the same notations and steps
used in Ref.~1 for the quark sector. Define four angular variables
$\theta_\downarrow,~\phi_\downarrow$ and
$\theta_\uparrow,~\phi_\uparrow$ by
$$
\xi_\downarrow=\tan \phi_\downarrow,~~\xi_\uparrow=\tan
\phi_\uparrow\eqno(1.20)
$$
$$
\eta_\downarrow=\tan \theta_\downarrow \cos \phi_\downarrow~~{\sf
and}~~\eta_\uparrow=\tan \theta_\uparrow \cos \phi_\uparrow.
\eqno(1.21)
$$
It is useful to introduce a subscript $s$, with
$$
s=~\downarrow~~{\sf or}~~\uparrow.\eqno(1.22)
$$
The eigenstates of the zeroth order mass operator $M_0(l_s)$ are
$$
\epsilon_s= \left(
\begin{array}{l}
\cos \theta_s\\
\sin \theta_s \cos \phi_s\\
\sin \theta_s \sin \phi_s
\end{array}
\right),
$$
$$
p_s = \left(
\begin{array}{l}
-\sin \theta_s\\
\cos \theta_s \cos \phi_s\\
\cos \theta_s \sin \phi_s
\end{array}
\right)
$$
and
$$
P_s = \left(
\begin{array}{l}
0\\
-\sin \phi_s\\
\cos \phi_s
\end{array}
\right)\eqno(1.23)
$$
with their corresponding eigenvalues given by
$$
\lambda(\epsilon_s)=0,\eqno(1.24)
$$
$$
\lambda(p_s)=\beta_s\bigg[1+\eta_s^2(1+\xi_s^2)\bigg]\eqno(1.25)
$$
and
$$
\lambda(P_s)=\alpha_s(1+\xi_s^2)+\beta_s.\eqno(1.26)
$$
Hence, these state-vectors are the zeroth order physical lepton
states with
$$
\epsilon_\downarrow =e,~~p_\downarrow=\mu,~~P_\downarrow=\tau,
$$
$$
\epsilon_\uparrow
=\nu_1,~~p_\uparrow=\nu_2,~~P_\uparrow=\nu_3.\eqno(1.27)
$$
From (1.23), one sees that the first elements of $P_\downarrow$ and
$P_\uparrow$ are both zero. (The motivation for imposing (1.18) is
essentially to achieve this fact which simplifies calculations.)
Hence, (1.19) follows.

Let $(V_s)_0$ be the $3\times 3$ unitary matrix that diagonalizes
the mass matrix $M_0(l_s)$ of (1.11)-(1.12). We have, by using
(1.22)-(1.24),
$$
(V_s)_0=(\epsilon_s~p_s~P_s).\eqno(1.28)
$$
The corresponding zeroth order $\nu$-mapping matrix is the $3\times
3$ real orthogonal matrix
$$
(V_{\nu-map})_0=(V_\downarrow^\dag)_0(V_\uparrow)_0.\eqno(1.29)
$$
By using (1.22)-(1.24), we find
$$
(V_{\nu-map})_0=\left(
\begin{array}{ccc}
\cos \theta_\downarrow\cos \theta_\uparrow &
-\sin\theta_\downarrow \cos\theta_\uparrow &\sin\theta_\uparrow \sin\phi\\
~~+\sin\theta_\downarrow \sin\theta_\uparrow \cos\phi
&~~+\cos\theta_\downarrow \sin\theta_\uparrow \cos\phi &\\
&&\\
-\cos \theta_\downarrow\sin \theta_\uparrow &
\sin\theta_\downarrow \sin\theta_\uparrow &\cos\theta_\uparrow \sin\phi\\
~~+\sin\theta_\downarrow \cos\theta_\uparrow \cos\phi
&~~+\cos\theta_\downarrow \cos\theta_\uparrow \cos\phi &\\
&&\\
 -\sin\theta_\downarrow \sin\phi
&-\cos\theta_\downarrow\sin\phi&\cos\phi
\end{array}
 \right )~,\eqno(1.30)
$$
in which $\phi_\downarrow$ and $\phi_\uparrow$ only appear through
their difference
$$
\phi=\phi_\uparrow-\phi_\downarrow.\eqno(1.31)
$$
Eqs. (1.19) and (1.31) account for the three "gauge" degrees of
freedom of (1.17).

The above description for the lepton sector corresponds exactly to
that for the quark sector in Ref.~1. Their difference lies only in
the magnitude of these three angles
$\theta_\downarrow,~\theta_\uparrow$ and $\phi$. For quarks, all
three angles are small. This is not the case for leptons. Using
Eq.(5.14) of Ref.~1, we have
$$
\xi_\downarrow = \tan \phi_\downarrow =-1
$$
$$
{\sf
and}~~~~~~~~~~~~~~~~~~~~~~~~~~~~~~~~~~~~~~~~~~~~~~~~~~~~~\eqno(1.32)
$$
$$
\eta_\uparrow=\tan \theta_\uparrow \cos \phi_\uparrow =
-\sqrt{\frac{1}{2}}
$$
which are both not small. On the other hand, the parameters
$$
x\equiv \xi_\uparrow=\tan \phi_\uparrow
$$
$$
{\sf
and}~~~~~~~~~~~~~~~~~~~~~~~~~~~~~~~~~~~~~~~~~~~~~~~~~\eqno(1.33)
$$
$$
y\equiv \eta_\downarrow=\tan \theta_\downarrow \cos
\phi_\downarrow
$$
are both small. Correspondingly, (1.11) and (1.12) become
$$
{\cal M}_0(l_\downarrow) = \alpha_\downarrow | l_3(\downarrow) +
l_2(\downarrow)|^2 + \beta_\downarrow | l_2(\downarrow) - y
l_1(\downarrow)|^2 +\beta_\downarrow |l_3(\downarrow) + y
l_1(\downarrow)|^2\eqno(1.34)
$$
and
$$
{\cal M}_0(l_\uparrow) = \alpha_\uparrow | l_3(\uparrow) - x
l_2(\uparrow)|^2 + \beta_\uparrow | l_2(\uparrow)
+\sqrt{\frac{1}{2}} l_1(\uparrow)|^2 +\beta_\uparrow |l_3(\uparrow)
+ \sqrt{\frac{1}{2}} x l_1(\uparrow)|^2. \eqno(1.35)
$$

When $x=y=0$, the neutrino mapping matrix $V_\nu$ takes on the
Harrison-Perkins-Scott form[5,6]. In that limit, $x=0$ implies
$\phi_\uparrow=0$ and therefore
$$
\hat{l}_3\parallel \hat{\nu}_3.\eqno(1.36)
$$
Likewise, when $y=0$, we have $\theta_\downarrow=0$ and
$$
\hat{l}_1 \parallel \hat{e}.\eqno(1.37)
$$
In section 2, we discuss the model with $T$ violation and evaluate
the masses of $e$ and $\nu_1$. The Jarlskog invariant of the
neutrino mapping matrix is calculated in section 3.


\section*{\Large \sf  2. $T$ violation}
\setcounter{section}{2} \setcounter{equation}{0}

With $T$ violation, we modify (1.34)-(1.35) by writing
$$
{\cal M}(l_\downarrow) = \alpha_\downarrow | l_3(\downarrow) +
e^{i\chi_\downarrow} l_2(\downarrow)|^2 + \beta_\downarrow |
l_2(\downarrow) - y l_1(\downarrow)|^2 +\beta_\downarrow
|l_3(\downarrow) + y l_1(\downarrow)|^2\eqno(2.1)
$$
and
$$
{\cal M}(l_\uparrow) = \alpha_\uparrow | l_3(\uparrow) - x
e^{i\chi_\uparrow} l_2(\uparrow)|^2 + \beta_\uparrow |
l_2(\uparrow) +\sqrt{\frac{1}{2}} l_1(\uparrow)|^2 +\beta_\uparrow
|l_3(\uparrow) + \sqrt{\frac{1}{2}} x l_1(\uparrow)|^2. \eqno(2.2)
$$
[Note that in Ref.~1, a different choice is made by placing the
$T$-violation factor $e^{i\chi_\uparrow}$ between $l_2(\uparrow)$
and $l_1(\uparrow)$.] By using (1.7)-(1.8), the determinants of the
corresponding mass matrices are
$$
|M (l_\downarrow)| =2 \alpha_\downarrow\beta_\downarrow^2 y^2
(1-\cos \chi_\downarrow) \eqno(2.3)
$$
and
$$
|M (l_\uparrow)| =\alpha_\uparrow\beta_\uparrow^2 x^2 (1-\cos
\chi_\uparrow). \eqno(2.4)
$$
The masses of $e$ and $\nu_1$ satisfy
$$
m_e(m_e-\lambda_\mu)(m_e-\lambda_\tau)=|M(l_\downarrow)|\eqno(2.5)
$$
and
$$
m_1(m_1-\lambda_2)(m_1-\lambda_3)=|M(l_\uparrow)|\eqno(2.6)
$$
where $\lambda_\mu,~\lambda_\tau,~\lambda_2$ and $\lambda_3$ are the
zeroth order masses of $\mu,~\tau,~\nu_2$ and $\nu_3$ given by
(1.25)-(1.27). Neglecting $m_e/m_\mu$ and $m_e/m_\tau$ corrections,
(2.5) leads to
$$
m_e m_\mu m_\tau \cong 2 \alpha_\downarrow \beta_\downarrow^2
y^2(1-\cos \chi_\downarrow).\eqno(2.7)
$$
Likewise, (1.25)-(1.27) and (1.32)-(1.33) give
$$
m_\mu \cong \beta_\downarrow(1+2y^2)\eqno(2.8)
$$
and
$$
m_\tau \cong 2 \alpha_\downarrow +\beta_\downarrow.\eqno(2.9)
$$
Assuming $m_1/m_2$ and $m_1/m_3$ are both small (for which there is
as yet no experimental evidence), (2.6) gives
$$
m_1 m_2 m_3 \cong \alpha_\uparrow\beta_\uparrow^2 x^2(1-\cos
\chi_\uparrow)\eqno(2.10)
$$
and (1.25)-(1.27), (1.32)-(1.33) give
$$
m_2 \cong \frac{1}{2}\beta_\uparrow(3+x^2)\eqno(2.11)
$$
and
$$
m_3\cong \alpha_\uparrow(1+x^2)+\beta_\uparrow.\eqno(2.12)
$$

\newpage

It is convenient to introduce in the $\downarrow$ sector a phase
transformation of its base-vectors:
$$
\left(
\begin{array}{l}
l_1(\downarrow)\\
l_2(\downarrow)\\
l_3(\downarrow)
\end{array}
\right)=\Omega_\downarrow \left(
\begin{array}{l}
e_1\\
e_2\\
e_3
\end{array}
\right) \eqno(2.13)
$$
where
$$
\Omega_\downarrow = \left(
\begin{array}{ccc}
1& 0 & 0\\
0 &e^{-i\frac{1}{2}\chi_\downarrow}&0\\
0 &0&e^{i\frac{1}{2}\chi_\downarrow}
\end{array}\right).\eqno(2.14)
$$
Hence (2.1) becomes
$$
{\cal
M}(l_\downarrow)=\alpha_\downarrow|e_3+e_2|^2+\beta_\downarrow|e_2
-ye^{i\chi_\downarrow/2}e_1|^2+\beta_\downarrow|e_3
+ye^{-i\chi_\downarrow/2}e_1|^2
$$
$$
=(\bar{e}_1~\bar{e}_2~\bar{e}_3)\bigg[H_0(\downarrow)
+H_1(\downarrow)+O(y^2)\bigg]
\left(
\begin{array}{l}
e_1\\
e_2\\
e_3
\end{array}
\right) \eqno(2.15)
$$
with
$$
H_0(\downarrow)=\left(
\begin{array}{ccc}
0& 0 & 0\\
0 &\alpha_\downarrow+\beta_\downarrow&\alpha_\downarrow\\
0 &\alpha_\downarrow&\alpha_\downarrow+\beta_\downarrow
\end{array}\right)\eqno(2.16)
$$
$$
H_1(\downarrow)=\beta_\downarrow y\left(
\begin{array}{ccc}
0& -e^{-i\frac{1}{2}\chi_\downarrow} & e^{i\frac{1}{2}\chi_\downarrow}\\
-e^{i\frac{1}{2}\chi_\downarrow} &0&0\\
e^{-i\frac{1}{2}\chi_\downarrow} &0&0
\end{array}\right).\eqno(2.17)
$$
Likewise, we write (2.2) as
$$
{\cal
M}(l_\uparrow)=\bigg(\bar{l}_1(\uparrow)~\bar{l}_2(\uparrow)
~\bar{l}_3(\uparrow)\bigg)
\bigg[H_0(\uparrow)+H_1(\uparrow)+O(x^2)\bigg] \left(
\begin{array}{l}
l_1(\uparrow)\\
l_2(\uparrow)\\
l_3(\uparrow)
\end{array}
\right) \eqno(2.18)
$$
where
$$
H_0(\uparrow)=\left(
\begin{array}{ccc}
\frac{1}{2}\beta_\uparrow& \sqrt{\frac{1}{2}}\beta_\uparrow & 0\\
\sqrt{\frac{1}{2}}\beta_\uparrow &\beta_\uparrow&0\\
0 &0&\alpha_\uparrow+\beta_\uparrow
\end{array}\right)\eqno(2.19)
$$
and
$$
H_1(\uparrow)=x \left(
\begin{array}{ccc}
0&0& \sqrt{\frac{1}{2}}\beta_\uparrow\\
0&0&-\alpha_\uparrow e^{-i\chi_\uparrow}\\
\sqrt{\frac{1}{2}}\beta_\uparrow &-\alpha_\uparrow
e^{i\chi_\uparrow}&0
\end{array}\right).\eqno(2.20)
$$
The matrices $H_0(\downarrow)$ and $H_0(\uparrow)$ can be readily
diagonalized:
$$
{\cal V}_\downarrow^\dag H_0(\downarrow){\cal V}_\downarrow
=\left(
\begin{array}{ccc}
0&0&0\\
0&m_0(\mu)&0\\
0&0&m_0(\tau)
\end{array}
\right )\eqno(2.21)
$$
$$
{\cal V}_\uparrow^\dag H_0(\uparrow){\cal V}_\uparrow =\left(
\begin{array}{ccc}
0&0&0\\
0&m_0(2)&0\\
0&0&m_0(3)
\end{array}
\right )\eqno(2.22)
$$
where
$$
m_0(\mu)=\beta_\downarrow,~~~m_0(\tau)=
2\alpha_\downarrow+\beta_\downarrow,\eqno(2.23)
$$
$$
m_0(2)=\frac{3}{2}\beta_\uparrow,~~~
m_0(3)=\alpha_\uparrow+\beta_\uparrow\eqno(2.24)
$$
and$$ {\cal V}_\downarrow=(\hat{\epsilon}~\hat{m}~\hat{t}),~~~{\cal
V}_\uparrow=(\hat{n}_1~\hat{n}_2~\hat{n}_3)\eqno(2.25)
$$
with
$$
\hat{\epsilon}=\left(
\begin{array}{l}
1\\
0\\
0
\end{array}
\right),~~~~\hat{m}=\sqrt{\frac{1}{2}}\left(
\begin{array}{r}
0\\
1\\
-1
\end{array}
\right),~~~~\hat{t}=\sqrt{\frac{1}{2}}\left(
\begin{array}{l}
0\\
1\\
1
\end{array}
\right)\eqno(2.26)
$$
and
$$
\hat{n}_1=\left(
\begin{array}{r}
\sqrt{\frac{2}{3}}\\
-\sqrt{\frac{1}{3}}\\
0
\end{array}
\right),~~~~\hat{n}_2=\left(
\begin{array}{r}
\sqrt{\frac{1}{3}}\\
\sqrt{\frac{2}{3}}\\
0
\end{array}
\right),~~~~\hat{n}_3=\left(
\begin{array}{l}
0\\
0\\
1
\end{array}
\right).\eqno(2.27)
$$
Correspondingly, define
$$
h_\downarrow \equiv {\cal V}_\downarrow^\dag H_1(\downarrow){\cal
V}_\downarrow~~{\sf and}~~h_\uparrow \equiv {\cal V}_\uparrow^\dag
H_1(\uparrow){\cal V}_\uparrow.\eqno(2.28)
$$
We find
$$
h_\downarrow=\sqrt{2}\beta_\downarrow y\left(
\begin{array}{ccc}
0& -\cos \frac{1}{2}\chi_\downarrow & i\sin\frac{1}{2}\chi_\downarrow\\
-\cos \frac{1}{2}\chi_\downarrow &0&0\\
-i\sin \frac{1}{2}\chi_\downarrow &0&0
\end{array}\right)\eqno(2.29)
$$
and
$$
h_\uparrow=x\left(
\begin{array}{ccc}
0& 0& A^*\\
0&0&B^*\\
A&B&0
\end{array}\right)\eqno(2.30)
$$
where
$$
A=\sqrt{\frac{1}{3}}(\beta_\uparrow+\alpha_\uparrow
e^{i\chi_\uparrow})\eqno(2.31)
$$
and
$$
B=\sqrt{\frac{1}{6}}(\beta_\uparrow-2\alpha_\uparrow
e^{i\chi_\uparrow}).\eqno(2.32)
$$
Represent $\hat{\epsilon},~\hat{m},~\hat{t}$ and
$\hat{n}_1,~\hat{n}_2,~\hat{n}_3$ of (2.26)-(2.27) by their
ket-vectors (in Dirac's notation)
$$
|\epsilon),~|m),~|t)
$$
$$
{\sf and}~~~~~~~~~~~~~~~~~~~~~~~~~~~~~~~~~\eqno(2.33)
$$
$$
|n_1),~|n_2),~|n_3).
$$
Correspondingly, we designate
$$
|e),~|\mu),~|\tau)
$$
$$
{\sf and}~~~~~~~~~~~~~~~~~~~~~~~~~~~~~~~~~\eqno(2.34)
$$
$$
|\nu_1),~|\nu_2),~|\nu_3)
$$
to be the physical lepton states. Introduce the transformation
matrices
$$
{\cal W}_\downarrow=\equiv \left(
\begin{array}{ccc}
(\epsilon|e)& (\epsilon|\mu) & (\epsilon|\tau)\\
(m|e) &(m|\mu)&(m|\tau)\\
(t|e) &(t|\mu)&(t|\tau)
\end{array}\right)\eqno(2.35)
$$
and
$$
{\cal W}_\uparrow=\equiv \left(
\begin{array}{ccc}
(n_1|\nu_1)& (n_1|\nu_2) & (n_1|\nu_3)\\
(n_2|\nu_1)& (n_2|\nu_2) & (n_2|\nu_3)\\
(n_3|\nu_1)& (n_3|\nu_2) & (n_3|\nu_3)
\end{array}\right).\eqno(2.36)
$$
To first order perturbation in $x$ and $y$, we find
$$
{\cal W}_\downarrow=1+\sqrt{2}\beta_\downarrow y\left(
\begin{array}{ccc}
0& -\frac{1}{m_0(\mu)}\cos \frac{1}{2}\chi_\downarrow
& \frac{i}{m_0(\tau)}\sin\frac{1}{2}\chi_\downarrow\\
\frac{1}{m_0(\mu)}\cos \frac{1}{2}\chi_\downarrow &0&0\\
 \frac{i}{m_0(\tau)}\sin \frac{1}{2}\chi_\downarrow &0&0
\end{array}\right)\eqno(2.37)
$$
and
$$
{\cal W}_\uparrow=1+x\left(
\begin{array}{ccc}
0& 0 & \frac{A^*}{\alpha_\uparrow+\beta_\uparrow}\\
0 &0&\frac{2B^*}{2\alpha_\uparrow-\beta_\uparrow}\\
-\frac{A}{\alpha_\uparrow+\beta_\uparrow}
&-\frac{2B}{2\alpha_\uparrow-\beta_\uparrow}&0
\end{array}\right).\eqno(2.38)
$$
Denote $U_\downarrow$ and $U_\uparrow$ to be the unitary matrices
that diagonalize the mass matrices $M(l_\downarrow)$ and
$M(l_\uparrow)$ defined by (1.7)-(1.8) and (2.1)-(2.2). To first
order in $x$ and $y$, we have
$$
U_\downarrow=\Omega_\downarrow{\cal V}_\downarrow{\cal
W}_\downarrow\eqno(2.39)
$$
and
$$
U_\uparrow={\cal V}_\uparrow{\cal W}_\uparrow.\eqno(2.40)
$$
Combining with $\Omega_\downarrow$ given by (2.14), ${\cal
V}_\downarrow,~{\cal V}_\uparrow$ by (2.25) and ${\cal
W}_\downarrow,~{\cal W}_\uparrow$ by (2.37)-(2.38), we derive
$$
U_\downarrow = \left(
\begin{array}{ccc}
1& -yX & iyY\\
\frac{yZ}{\sqrt{2}}e^{-i\frac{1}{2}\chi_\downarrow}
&\frac{1}{\sqrt{2}}e^{-i\frac{1}{2}\chi_\downarrow}
&\frac{1}{\sqrt{2}}e^{-i\frac{1}{2}\chi_\downarrow}\\
-\frac{yZ^*}{\sqrt{2}}e^{i\frac{1}{2}\chi_\downarrow}
&-\frac{1}{\sqrt{2}}e^{i\frac{1}{2}\chi_\downarrow}
&\frac{1}{\sqrt{2}}e^{i\frac{1}{2}\chi_\downarrow}
\end{array}\right)\eqno(2.41)
$$
with
$$
X=\sqrt{2}\cos \frac{1}{2}\chi_\downarrow,~~
Y=\frac{\sqrt{2}\beta_\downarrow}{2\alpha_\downarrow+\beta_\downarrow}
\sin \frac{1}{2}\chi_\downarrow\eqno(2.42)
$$
and
$$
Z=X+iY\eqno(2.43).
$$
Likewise,
$$
U_\uparrow = \left(
\begin{array}{ccc}
\sqrt{\frac{2}{3}}& \sqrt{\frac{1}{3}} &x\sqrt{\frac{1}{3}}
(\frac{\sqrt{2}A^*}{\alpha_\uparrow+\beta_\uparrow}+\frac{2 B^*}{2
\alpha_\uparrow-\beta_\uparrow})\\
 - \sqrt{\frac{1}{3}}&\sqrt{\frac{2}{3}}
&x\sqrt{\frac{1}{3}}
(\frac{-A^*}{\alpha_\uparrow+\beta_\uparrow}+\frac{2 \sqrt{2}B^*}{2
\alpha_\uparrow-\beta_\uparrow})\\
-\frac{xA}{\alpha_\uparrow+\beta_\uparrow} &-\frac{2xB}{2
\alpha_\uparrow-\beta_\uparrow} &1
\end{array}\right)\eqno(2.44)
$$
with $A$ and $B$ given by (2.31) and (2.32). The neutrino mapping
matrix $U_{\nu-map}$ is then related to (2.41) and (2.44) by
$$
U_{\nu-map}=U_\downarrow^\dag U_\uparrow.\eqno(2.45)
$$
Note that in accordance with our definitions (2.39)-(2.40)
$U_\uparrow$ and $U_\downarrow$ refer to the transformation matrices
relating the "bare"  annihilation operators
$l_i(\downarrow),~l_i(\uparrow)$ defined by the mass operators
(1.7)-(1.8) to the corresponding "physical" annihilation operators
of leptons (not their state vectors). Hence, the $\nu$-mapping
matrix in the particle data group literature is $U_{\nu-map}^*$, the
complex conjugate of (2.45).


\section*{\Large \sf  3. Jarlskog Invariant}
\setcounter{section}{3} \setcounter{equation}{0}

The matrix $U_{\nu-map}$ can be written as
$$
U_{\nu-map} = \left(
\begin{array}{ccc}
U_{e1}& U_{e2} &U_{e3}\\
U_{\mu 1}& U_{\mu 2} &U_{\mu 3}\\
U_{\tau 1}& U_{\tau 2} &U_{\tau 3}
\end{array}\right).\eqno(3.1)
$$
From (2.41), (2.44)-(2.45) and to first order in $x$ and $y$, we
find
$$
U_{e1}=\sqrt{\frac{2}{3}}-y\sqrt{\frac{1}{6}}
Z^*e^{i\frac{1}{2}\chi_\downarrow}
$$
$$
U_{e2}=\sqrt{\frac{1}{3}}+y\sqrt{\frac{1}{3}}
Z^*e^{i\frac{1}{2}\chi_\downarrow}
$$
$$
U_{e3}=x\sqrt{\frac{1}{3}}\bigg(\frac{\sqrt{2}
A^*}{\alpha_\uparrow+\beta_\uparrow}+\frac{2B^*}
{2\alpha_\uparrow-\beta_\uparrow}\bigg) -y\sqrt{\frac{1}{2}} Z
e^{-i\frac{1}{2}\chi_\downarrow}
$$
$$
U_{\mu
1}=-y\sqrt{\frac{2}{3}}X-\sqrt{\frac{1}{6}}e^{i\frac{1}{2}\chi_\downarrow}
+x\sqrt{\frac{1}{2}}
\frac{A}{\alpha_\uparrow+\beta_\uparrow}e^{-i\frac{1}{2}\chi_\downarrow}
$$
$$
U_{\mu
2}=-y\sqrt{\frac{1}{3}}X+\sqrt{\frac{1}{3}}e^{i\frac{1}{2}\chi_\downarrow}
+x\frac{\sqrt{2}B}{2\alpha_\uparrow-\beta_\uparrow}
e^{-i\frac{1}{2}\chi_\downarrow}\eqno(3.2)
$$
$$
U_{\mu 3}=x\sqrt{\frac{1}{6}}\bigg(-\frac{
A^*}{\alpha_\uparrow+\beta_\uparrow}+\frac{2
\sqrt{2}B^*}{2\alpha_\uparrow-\beta_\uparrow}\bigg)
e^{i\frac{1}{2}\chi_\downarrow} -\sqrt{\frac{1}{2}}
e^{-i\frac{1}{2}\chi_\downarrow}
$$
$$
U_{\tau
1}=-iy\sqrt{\frac{2}{3}}Y-\sqrt{\frac{1}{6}}e^{i\frac{1}{2}\chi_\downarrow}
-x\sqrt{\frac{1}{2}}
\frac{A}{\alpha_\uparrow+\beta_\uparrow}e^{-i\frac{1}{2}\chi_\downarrow}
$$
$$
U_{\tau
2}=-iy\sqrt{\frac{1}{3}}Y+\sqrt{\frac{1}{3}}e^{i\frac{1}{2}\chi_\downarrow}
-x\frac{\sqrt{2}B}{2\alpha_\uparrow-\beta_\uparrow}e^{-i\frac{1}{2}\chi_\downarrow}
$$
and
$$
U_{\tau 3}=x\sqrt{\frac{1}{6}}\bigg(-\frac{
A^*}{\alpha_\uparrow+\beta_\uparrow}+\frac{2
\sqrt{2}B^*}{2\alpha_\uparrow-\beta_\uparrow}\bigg)
e^{i\frac{1}{2}\chi_\downarrow} +\sqrt{\frac{1}{2}}
e^{-i\frac{1}{2}\chi_\downarrow}
$$
where $A,~B,~X,~Y$ and $Z$ are given by (2.31)-(2.32) and
(2.42)-(2.43).

Define
$$
T_1=U_{e1}^* U_{\mu 1},~~T_2=U_{e2}^* U_{\mu 2}~~{\sf
and}~~T_3=U_{e3}^* U_{\mu 3}.\eqno(3.3)
$$
By using (3.2), we have
$$
T_1=-\frac{1}{3}e^{i\frac{1}{2}\chi_\downarrow}-y\frac{2}{3}X
+x\sqrt{\frac{1}{3}}\frac{A}{\alpha_\uparrow+\beta_\uparrow}
e^{-i\frac{1}{2}\chi_\downarrow}+y\frac{1}{6}Z,
$$
$$
T_2=~~\frac{1}{3}e^{i\frac{1}{2}\chi_\downarrow}-y\frac{1}{3}X+x\sqrt{\frac{2}{3}}
\frac{B}{2\alpha_\uparrow-\beta_\uparrow}
e^{-i\frac{1}{2}\chi_\downarrow}+y\frac{1}{3}Z\eqno(3.4)
$$
and
$$
T_3=-x\sqrt{\frac{1}{3}}\bigg(\frac{
A}{\alpha_\uparrow+\beta_\uparrow}+\frac{
\sqrt{2}B}{2\alpha_\uparrow-\beta_\uparrow}\bigg)
e^{-i\frac{1}{2}\chi_\downarrow} +y\frac{1}{2}Z^*  ~~.
$$
Thus,
$$
T_1+T_2+T_3=0~~.\eqno(3.5)
$$
The Jarlskog invariant $J_{\nu-map}$ for the neutrino mapping
matrix is given by $Im T_1^*T_2$. We find
$$
J_{\nu-map}=-\frac{y}{6\sqrt{2}}
\bigg(1+\frac{\beta_\downarrow}{2\alpha_\downarrow+\beta_\downarrow}\bigg)
~\sin\chi_\downarrow
+\frac{\alpha_\uparrow \beta_\uparrow
x}{3(\alpha_\uparrow+\beta_\uparrow)
(2\alpha_\uparrow-\beta_\uparrow)}\bigg[
\sin\chi_\downarrow+\sin(\chi_\uparrow-\chi_\downarrow)\bigg]\eqno(3.6)
$$
which is valid for small $x$ and $y$. If in addition
$\chi_\downarrow$ and $\chi_\uparrow$ are also small, then
$$
J_{\nu-map}\cong -\frac{y \chi_\downarrow}{6\sqrt{2}}
\bigg(1+\frac{\beta_\downarrow}{2\alpha_\downarrow+\beta_\downarrow}\bigg)
+\frac{x \chi_\uparrow \alpha_\uparrow \beta_\uparrow
}{3(\alpha_\uparrow+\beta_\uparrow)
(2\alpha_\uparrow-\beta_\uparrow)}~.\eqno(3.7)
$$
From (2.7) and (2.23) we find
$$
m_\mu\cong \beta_\downarrow,~~~m_\tau\cong
2\alpha_\downarrow+\beta_\downarrow
$$
$$
{\sf
and}~~~~~~~~~~~~~~~~~~~~~~~~~~~~~~~~~~~~~~~~~~~~~~~~~~~~~~~~~~\eqno(3.8)
$$
$$
m_e m_\mu m_\tau \cong \alpha_\downarrow \beta_\downarrow^2 y^2
\chi_\downarrow^2.
$$
Thus,
$$
y\chi_\downarrow\cong
\pm\sqrt{\frac{2m_em_\tau}{m_\mu(m_\tau-m_\mu)}}.\eqno(3.9)
$$
Likewise, from (2.10) and (2.24)
$$
m_2\cong \frac{3}{2}\beta_\uparrow,~~m_3\cong
\alpha_\uparrow+\beta_\uparrow
$$
$$
{\sf
and}~~~~~~~~~~~~~~~~~~~~~~~~~~~~~~~~~~~~~~~~~~~~~~~~~~~~~~~~\eqno(3.10)
$$
$$
m_1m_2m_3\cong
\frac{1}{2}\alpha_\uparrow\beta_\uparrow^2x^2\chi_\uparrow^2,
$$
which lead  to
$$
x\chi_\uparrow\cong
\pm3\sqrt{\frac{3m_1m_3}{2m_2(3m_3-2m_2)}}.\eqno(3.11)
$$
Write (3.7) as
$$
J_{\nu-map}=J_e+J_\nu\eqno(3.12)
$$
in which
$$
J_e=\frac{1}{6}\sqrt{\frac{m_e}{m_\mu}}
\bigg[\frac{m_\tau+m_\mu}{\sqrt{m_\tau(m_\tau-m_\mu)}}\bigg]
$$
$$
=\frac{1}{6}\bigg[\sqrt{\frac{m_e}{m_\mu}}
+O\bigg(\sqrt{\frac{m_em_\mu}{m_\tau^2}}\bigg)\bigg]\eqno(3.13)
$$
and if $m_1/m_2$ and $m_2/m_3$ are both $<<1$ then
$$
J_\nu\cong
\pm\frac{1}{3\sqrt{2}}\sqrt{\frac{m_1m_2}{m_3^2}}.\eqno(3.14)
$$
For convenience, we set the sign of $J_e$ to be positive, thus,
$$
J_{\nu-map}\cong \frac{1}{6}\sqrt{\frac{m_e}{m_\mu}}
\bigg[\frac{m_\tau+m_\mu}{\sqrt{m_\tau(m_\tau-m_\mu)}}\bigg] \pm
\frac{1}{3\sqrt{2}}\sqrt{\frac{m_1m_2}{m_3^2}}\eqno(3.15)
$$
which leads to (1.2).

\newpage

\section*{\Large \sf Appendix}

Consider the eigenstate equation
$$
H|i)=m_i|i)\eqno(A.1)
$$
of a $3\times 3$ hermitian Hamiltonian
$$
H=H_0+h\eqno(A.2)
$$
with $i=1,~2,~3$ and $H_0$ diagonal. In the case that the
eigenvalues $m_i$ are known (or can be derived by a simple series
expansion as in the case of the masses of $e,~\mu,~\tau$ and
$\nu_1,~\nu_2,~\nu_3$ through (2.5)-(2.6)), the explicit form of
the eigenvectors $|i)$ can be obtained by solving a simple linear
equation of two variables, as we shall see. Write
$$
H_0 = \left(
\begin{array}{ccc}
n_\xi& 0 & 0\\
0 &n_\eta&0\\
0 &0&n_\zeta
\end{array}\right)\eqno(A.3)
$$
and
$$
h = \left(
\begin{array}{ccc}
h_{\xi\xi}& h_{\xi\eta} & h_{\xi\zeta}\\
h_{\eta\xi} &h_{\eta\eta}&h_{\eta\zeta}\\
h_{\zeta\xi} &h_{\zeta\eta}&h_{\zeta\zeta}
\end{array}\right).\eqno(A.4)
$$
Denote the normalized eigenstates of $H_0$ by $|s)$, so that
$$
H_0|s)=n_s|s).\eqno(A.5)
$$
with $s=\xi,~\eta,~\zeta$ and the normalization
$$
(\xi|\xi)=(\eta|\eta)=(\zeta|\zeta)=1.\eqno(A.6)
$$
However, for the eigenstates of $H$ we choose the Brillouin-Wigner
normalization condition with
$$
(\xi|1)=(\eta|2)=(\zeta|3)=1.\eqno(A.7)
$$
The three eigenstates $|i)$ of (A.1) can then be written as
$$
|1)=|\xi)+(\eta|1)|\eta)+(\zeta|1)|\zeta)
$$
$$
|2)=(\xi|2)|\xi)+|\eta)+(\zeta|2)|\zeta)\eqno(A.8)
$$
$$
|3)=(\xi|3)|\xi)+(\eta|3)|\eta)+|\zeta).
$$

Since
$$
H|1)=m_1|1)\eqno(A.9)
$$
we have
$$
(\xi|H|1)=m_1(\xi|1)=m_1.\eqno(A.10)
$$
On the other hand,
$$
H|1)=H_0|1)+h|1)\eqno(A.11)
$$
which gives
$$
(\xi|H|1)=n_\xi(\xi|1)+(\xi|h|1)=n_\xi+(\xi|h|1).\eqno(A.12)
$$
Thus, (A.10) and (A.12) yield
$$
m_1=n_\xi+(\xi|h|1).\eqno(A.13)
$$
Likewise, from (A.9) we find
$$
(\eta|H|1)=m_\eta(\eta|1)\eqno(A.14)
$$
and by using (A.11)
$$
(\eta|H|1)=n_\eta(\eta|1)+(\eta|h|1).\eqno(A.15)
$$
Combining (A.14) and (A.15), we derive
$$
(\eta|1)=\frac{1}{m_1-m_\eta}(\eta|h|1)\eqno(A.16)
$$
and, in identical way,
$$
(\zeta|1)=\frac{1}{m_1-m_\zeta}(\zeta|h|1).\eqno(A.17)
$$

Next, introduce
$$
x=(\eta|1),~~y=(\zeta|1)\eqno(A.18)
$$
and write the first equation in (A.8) as
$$
|1)=|\xi)+x|\eta)+y|\zeta).\eqno(A.19)
$$
This leads to
$$
h|1)=h|\xi)+xh|\eta)+yh|\zeta);\eqno(A.20)
$$
therefore
$$
(\eta|h|1)=h_{\eta\xi}+xh_{\eta\eta}+yh_{\eta\zeta}\eqno(A.21)
$$
and
$$
(\zeta|h|1)=h_{\zeta\xi}+xh_{\zeta\eta}+yh_{\zeta\zeta}.\eqno(A.22)
$$
From (A.16), (A.18) and (A.21) it follows
$$
x=\frac{1}{m_1-m_\eta}[h_{\eta\xi}+xh_{\eta\eta}+yh_{\eta\zeta}].\eqno(A.23)
$$
Likewise, from (A.17) and (A.22),
$$
y=\frac{1}{m_1-m_\zeta}[h_{\zeta\xi}+xh_{\zeta\eta}+yh_{\zeta\zeta}].\eqno(A.24)
$$
Define
$$
\Delta(1)=(m_1-n_\eta-h_{\eta\eta})(m_1-n_\zeta
-h_{\zeta\zeta})-h_{\eta\zeta}h_{\zeta\eta}.\eqno(A.25)
$$
By using (A.18) and (A.23)-(A.24), we find
$$
(\eta|1)=\frac{1}{\Delta(1)}\bigg[(m_1-n_\zeta-h_{\zeta\zeta})h_{\eta\xi}
+h_{\eta\zeta}h_{\zeta\xi}\bigg]\eqno(A.26)
$$
and
$$
(\zeta|1)=\frac{1}{\Delta(1)}\bigg[(m_1-n_\eta-h_{\eta\eta})h_{\zeta\xi}
+h_{\zeta\eta}h_{\eta\xi}\bigg].\eqno(A.27)
$$
Thus we have the explicit solution of $|1)$, with the
Brillouin-Wigner normalization introduced in (A.8). Likewise, the
other two eigenstates $|2)$ and $|3)$ can be similarly derived. We
find
$$
(\zeta|2)=\frac{1}{\Delta(2)}\bigg[(m_2-n_\xi-h_{\xi\xi})h_{\zeta\eta}
+h_{\zeta\xi}h_{\xi\eta}\bigg],\eqno(A.28)
$$
$$
(\xi|2)=\frac{1}{\Delta(2)}\bigg[(m_2-n_\zeta-h_{\zeta\zeta})h_{\xi\eta}
+h_{\xi\zeta}h_{\zeta\eta}\bigg],\eqno(A.29)
$$
$$
(\xi|3)=\frac{1}{\Delta(3)}\bigg[(m_3-n_\eta-h_{\eta\eta})h_{\xi\zeta}
+h_{\xi\eta}h_{\eta\zeta}\bigg]\eqno(A.30)
$$
and
$$
(\eta|3)=\frac{1}{\Delta(3)}\bigg[(m_3-n_\xi-h_{\xi\xi})h_{\eta\zeta}
+h_{\eta\xi}h_{\xi\zeta}\bigg]\eqno(A.31)
$$
with
$$
\Delta(2)=(m_2-n_\zeta-h_{\zeta\zeta})(m_2-n_\xi
-h_{\xi\xi})-h_{\zeta\xi}h_{\xi\zeta}\eqno(A.32)
$$
and
$$
\Delta(3)=(m_3-n_\xi-h_{\xi\xi})(m_3-n_\eta
-h_{\eta\eta})-h_{\xi\eta}h_{\eta\xi}.\eqno(A.33)
$$
These formulas would be useful for higher order corrections to
$U_{\nu-map}$ and $J_{\nu-map}$.

\section*{\Large \sf References}


\noindent [1] R. Friedberg and T. D. Lee, Ann. Phys. in press\\

\noindent [2] S. Eidelman et al. (Particle Data Group), Phys.
Lett. B592, 1(2004).\\

\noindent [3] T. D. Lee, Phys. Rev. D8, 1226(1973)\\

\noindent [4] T. D. Lee, Physics Reports 9c, 2(1974)\\

\noindent [5] P. F. Harrison, D. H. Perkins and W. G. Scott,

~~~~~~~~~~Phys. Lett. B530, 167(2002).\\

\noindent [6] L. Wolfenstein, Phys. Rev. D18, 958(1978);

P. F. Harrison and W. G. Scott, Phys. Lett. B535, 163(2002);

Z. Z. Xing, Phys. Lett. B533, 85(2002);

X. G. He and A. Zee, Phys. Lett. B560, 87(2003).\\

\newpage

\end{document}